\documentclass[12pt]{article}

\usepackage{epsfig}
\usepackage[usenames,dvipsnames]{color}
\usepackage{float}

\newcommand{\ai}{\'{\i}}
  
\begin{document}

\begin{center}
{\Large{\bf Nonsingular Cosmological Models}}
\end{center}
\begin{center}
\large{Santiago Esteban Perez Bergliaffa}\footnote{Email: sepbergliaffa@gmail.com}\\
\normalsize{\textsl{ Departamento de F\ai sica Te\'orica,
Instituto de F\ai sica, \\Universidade do Estado do Rio de Janeiro}}\\
\normalsize{\textsl{Rua S\~ao Francisco Xavier 524, Maracan\~a,
Rio de Janeiro, Brasil,\\
CEP: 20550-900 .}}
\end{center}




\begin{abstract}
A short introduction to cosmological models 
that go 
from
an era of accelerated collapse to an expanding era without displaying 
a singularity is presented. 
\end{abstract}




\section{Introduction}
\label{sec:intro}

The standard cosmological model (SCM) (see for instance 
\cite{pdg}
for an updated review) furnishes
an accurate and succesful description of the evolution of the universe, which spans approximately 14
billion years. The main hypothesis on which the model is based are the following:
\begin{itemize}
\item Gravity is described by General Relativity,
\item The Cosmological Principle, 
\item Above a certain scale, the matter content of the model is described by a continuous
distribution of matter/energy, which is described by a perfect fluid.
\end{itemize}
In spite of its success, the SCM suffers from a series of problems such as the initial singularity,
the cosmological horizon, the flatness problem, the baryon asymmetry, and the nature of dark
matter and dark energy \footnote{Some "open questions" may be added to this list, such as why the Weyl tensor is null, and what the
future evolution of the universe is.}. 
Although inflation (which for many is currently a part of the
SCM) partially or totally answers some of these, it does not solve the crucial issue of the
initial singularity
\footnote{In fact, inflation presents some problems of its own, such as the identification of the inflaton with a definite
field of some high-energy theory, the functional form of the potential in terms of the inflaton, and the transplanckian problem. See for instance \cite{braninfl}.}. 

The existence of an initial singularity is disturbing: a singularity can
be naturally considered as a source of lawlessness \footnote{For a discussion of the singularity theorems and of the concept of singularity see for instance
\cite{earman}).}, 
because our description of spacetime
breaks down ``there'', and physical laws presuppose spacetime. Regardless of the fact that
several scenarios have been developed to deal with the singularity issue, the breakdown of
physical laws continues to be a conundrum after almost a hundred years of the discovery
of the FLRW solution \footnote{This acronym refers to the authors that presented for the first time the solution of EE that describes
a universe with zero pressure (Friedmann) and nonzero pressure (Lem\^{a}itre), and to those who
studied its general mathematical properties and took it to its current form (Robertson and Walker). For historical details, see \cite{merlau}.}
(which inevitably displays a past singularity).

The initial singularity is distressing for many other reasons
\footnote{See \cite{nossoreport} for a complete list, as well as 
a detailed revision of nonsingular cosmological models.}.
To name just two, the Cauchy problem is not well-formulated in spacetimes with a singularity, 
and the initial singularity is inconsistent with the entropy bound \cite{ebound}. There are also hints that 
quantum gravitational effects may tame the singularity, as a consequence of the
discreteness of the spectrum of some operators.
As a consequence of all these arguments indicating that the initial singularity may be absent
in realistic descriptions of the universe, many cosmological solutions displaying a bounce
were examined in the last decades, starting from
the first explicit
solutions for a bouncing geometry obtained by \cite{novsal} and \cite{melni}.
In fact, 
there is a "window of
opportunity" to avoid the initial singularity in FLRW models at a classical level by one or a
combination of the following assumptions:
\begin{itemize}
\item Violating strong energy condition in the realm of GR;
\item Working with a new gravitational theory, as for instance those that add scalar degrees
of freedom to gravity (Brans-Dicke theory being the paradigmatic example of this type), or by adopting an action built with higher-order invariants.
\end{itemize}
Other ways to avoid the singularity are:
\begin{itemize}
\item Changing the way gravity couples to matter, from minimal to non-minimal coupling;
\item Using a non-perfect fluid as a source.
\end{itemize}
Finally, quantum gravitational effects also give the chance of a bounce
\footnote{See \cite{nossoreport} for details about all these items.}.
In the next section we shall briefly discuss 
how a bounce can solve some of the problems that the cosmological model 
pre-1980 had.

\section{The bounce and the problems of the standard cosmology}

\begin{figure}
       \centering  
       \includegraphics[scale=.60]{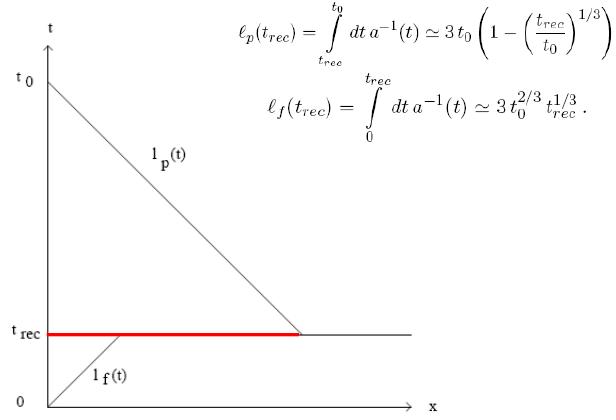}
  \caption{The plot of the physical distance $x$ versus time $t$ illustrates the
homogeneity problem: the past light cone $\ell_p(t)$ at the time $t_{rec}$ (in red)
is much
larger than the forward light cone $\ell_f (t)$ at $t_{rec}$. Adapted from \cite{branfig}.}
  \label{ho}     
\end{figure}

\begin{figure}
       \centering  
       \includegraphics[scale=.80]{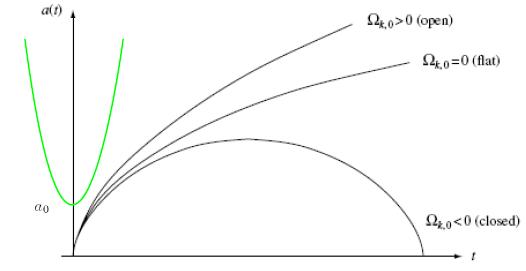}
  \caption{Typical evolution of the scale factor in a nonsingular cosmological model (in green),
  as opposed to the singular big bang model. After the bounce, the evolution must enter the radiation era 
  timely.}
  \label{bou}
\end{figure}

In addition to the initial singularity, the
SCM had 
other problems. Among them we can cite the following
\cite{branprob}:
\begin{itemize}
\item The homogeneity problem: the comoving
region over which the CMB is observed to be homogeneous
to better than one part in $10^{-4}$ is much larger than the comoving forward
light cone at the time of recombination (see figure \ref{ho}).
\item The flatness problem: the quantity $|\Omega -1|$ decreases with the evolution
of a universe dominated by matter or radiation. Since $\Omega \approx 1$ today \cite{komatsu}, 
$\Omega$ must have been incredibly close to 1 in the past.
\item The generation of primordial perturbations: clusters of galaxies have nonrandom correlations on scales larger than
50 Mpc. This scale is comparable to the comoving horizon
at $t_{eq}$. If the initial density perturbations were produced much before
$t_{eq}$, the correlations cannot be explained by a causal mechanism
\footnote{Actually,  standard cosmology cannot explain how primordial density perturbations
are generated.}.
\end{itemize}
Except for the inital singularity, 
these problems were addressed by inflation (which has problems of its own
as we mentioned before) \footnote{For a review of inflation, see for instance
\cite{bau}.}. 
We shall see next that a model with a bounce may also face these issues 
succesfully.
Let us state that by a nonsingular model with a LFRW geometry we mean a model in 
which the scale factor attains a minimum value (figure \ref{bou})
Consequently, a model with a bounce solves the problem of the initial singularity by construction. 
Regarding the homogeneity problem, 
the future light cone is given by
$$
{\ell}_f(t)=a(t)\int_{t_i}^t \frac{dt}{a(t)}. 
$$
Assuming the equation of state
$
p=\omega \rho$
it follows that
$
a(t) \propto (-t)^{\frac{2}{3(1+\omega )}}.$
Hence,
$$
{\ell}_f(t) \propto (-t_i)^{\frac{1+3\omega}{3(1+\omega)}} (-t)^{\frac{2}{3(1+\omega )}}+t
$$
If there is a contracting phase led by a perfect fluid with
$
\omega > -1/3 
$, then 
${\ell}_f(t)$ diverges for
$t_i\rightarrow\infty$, thus solving the horizon problem. 
The flatness problem
is encoded in the equation
$$
\frac{d}{dt}|\Omega - 1| = -2\frac{\ddot a}{\dot a^3}.
$$
Since the standard evolution drives $\Omega$ to 1, 
an era during which the evolution of the universe 
forces $|\Omega - 1|$ away of zero
is needed. 
This can achieved by an expansion such that 
$\ddot a>0$ and $\dot a>0$ (which is the case 
of inflation), or through a long decelerated phase of
contraction before the bounce, characterized
 by $\ddot a <0$ and $\dot a<0$. 

\begin{figure}
       \centering  
       \includegraphics[scale=.60]{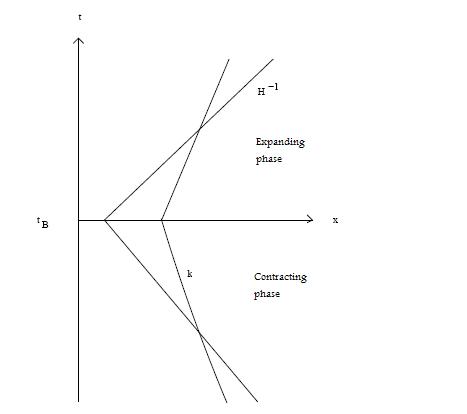}
  \caption{The plot shows the evolution of the Hubble radius $H^{-1}$
and of a fixed comoving scale, the bounce taking place at $t=t_B$. Adapted from \cite{branfig}.}
  \label{gp}
\end{figure}

Regarding the generation of primordial perturbations in nonsingular models,
during the contracting phase the Hubble radius $H^{-1}$ contracts
faster than the physical length corresponding to a fixed comoving scale $k$
(see
figure \ref{gp}). Quantum vacuum
fluctuations generated causally on sub-Hubble scales in the contracting phase
are assumed to be the seeds of the inhomogeneities observed today. The 
scale of these 
fluctuations is amplified and evolves according to GR during the (long) time when it is 
larger than the Hubble radius. 
Finally, if the bounce is such that 
$
a_0>> {\ell}_{\rm Pl}
$
there is no transplanckian problem.
\section{An example}
\label{ex}
Having shown in the previous section that a nonsingular model may 
in principle furnish a solution to the problems of standard cosmology, 
let us review in this section a specific model of this type, and what
kind of predictions can be obtained from it. The model in question
was developed in \cite{nelsonpatrick}, in the framework of GR plus 
a perfect fluid with equation of state $p=\omega \rho$, the spacetime
geometry 
being of the FLRW type. The quantization of both the background and the perturbations of this model
following the Bohmian approach (see \cite{nelsonpatrick} for details),
furnishes for the evolution of the background
$$
a(\tau ) = a_0\left[1+\left(\frac{\tau}{T_0}\right)^2\right]^{1/[3(1-\omega)]},
$$
with 
$d\eta = [a(\tau )]^{3\omega -1}d\tau$, and $\eta$ is the conformal time.
This solution has no singularities and tends
to the classical solution when $\tau\rightarrow\pm\infty$.
An analysis of the perturbations shows that they behave exactly as
shown in figure \ref{gp}. The result obtained in \cite{nelsonpatrick}
for the power spectra is
$$
n_s=1+\frac{12\omega}{1+3\omega},\;\;\;\;\;n_T=\frac{12\omega}{1+3\omega},
$$
in such a way that both the scalar and the tensor spectrum tend to a scale-invariant
spectrum in the dust limit. Finally, a fit of the amplitude of the perturbations to the 
CMB data yields $a_0\approx 1000 {\ell_{\rm Planck}}$, thus avoiding the transplanckian problem.
Notice also that the model predicts a tensor to scalar ratio of
$T/S\propto \sqrt{n_s-1}$, while inflationary models tipically predict a linear relation.

\section{Bouncing models and observation}

\begin{figure}
       \centering  
       \includegraphics[scale=.80]{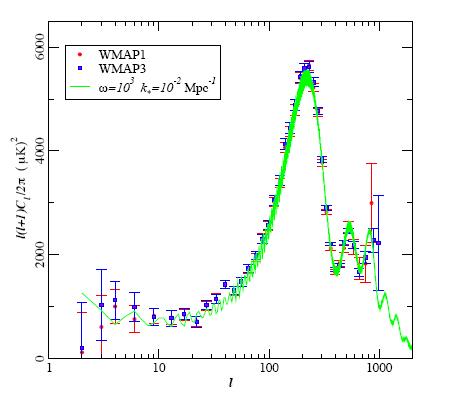}
       \caption{The plot shows the multipoles $C_\ell$ for a typical bouncing model, as well as 
  WMAP data \cite{falciano}.}
  \label{osci}
\end{figure}

As we have seen in the previous sections, nonsingular models may solve 
the problems of standard cosmology. The example discussed in Sec.\ref{ex}
shows that some particular models produce predictions that are 
not incompatible with observations. Some other predictions generic to 
bouncing models are:
\begin{itemize}
\item The spectrum of primordial perturbations displays a small oscillatory component,
see figure \ref{osci}
\cite{falciano}.
\item Copious production of particles near the bounce. This has been estimated 
in the case of gravitons in the Pre-Big Bag model \cite{vene}, and for photons 
in the WIST theory \cite{nilton1},\cite{nilton2}.
\end{itemize}
\section{Conclusion}
Nonsingular models offer the chance of solving the problems of standard cosmology,
and furnish predictions that may be contrasted with
observation in the near future. 
There are still issues to be solved (such as
the influence on the perturbations of the matter creation at the bounce,
the amount of matter created, and the possible growth of 
initial perturbations), but the bottom line is that 
models with a bounce are certainly worth studying,
on their own sake and/or as a complement to inflation
\footnote{See for instance \cite{cai}.}.
\section*{Acknowledgements}
The author would like to acknowledge financial support from UERJ, FAPERJ
and CNPQ.

\end{document}